\documentclass[12pt]{article}
\usepackage{latexsym}
\begin{document}
\title{Note on the energy-momentum tensor for general mixed tensor-spinor fields}
\author{\begin{tabular}{c}\bigskip Hongbao Zhang\footnote{Email: hbzhang@pkuaa.edu.cn}\\  \smallskip Department of Physics, Beijing Normal University, Beijing, 100875, PRC
\end{tabular}}
\maketitle
\begin{abstract}
This note provides an explicit proof of the equivalence of the
Belinfante's energy-momentum tensor and the metric energy-momentum
tensor for general mixed tensor-spinor fields.
\end{abstract}
\section{Introduction}
It is well known that a suitable definition for the
energy-momentum tensor has been under investigation for many
years. We require $T_{ab}$ constructed should not only provide
meaningful physical conserved quantities but also act as the
source of Einstein's gravitational field equation.

In a flat spacetime, the canonical energy-momentum tensor arises
from Noether's Theorem by employing the conserved currents
associated with translational symmetries\cite{Wald}. However, only
for the scalar field the canonical energy-momentum tensor turns
out to be symmetric, satisfying both of the requirements above.
Indeed, in the case of the Maxwell field, the canonical
energy-momentum tensor is not even gauge invariant\cite{Jack}. Of
course, for general higher spin fields, we can correct it through
Belinfante's symmetrization procedure\cite{Belin}, although this
is usually presented as an {\em ad hoc} prescription\cite{Weinb1}.

On the other hand, a completely different approach, based on the
powerful variational principle, leads to the metric
energy-momentum tensor, which naturally satisfies our requirements
above\cite{Wald,Weinb2}.

It is certainly interesting to ask if the Belinfante's
energy-momentum tensor is equivalent to the metric energy-momentum
tensor. Despite a general belief in their equivalence, to our
knowledge there exists no such an explicit proof in the literature
except that \cite{Sara} presents a detailed confirmation for
general tensor fields. The aim of this note is to generalize this
confirmation to the case of general mixed tensor-spinor fields.
Accordingly, the present work acquires its practical importance:
it justifies much work such as \cite{Vollick,YW} and references
therein, where the Belinfante's energy momentum tensor for fermion
fields is imposed as the source of Einstein's gravitation.

Notation and conventions adopted by Chapter 13 in \cite{Wald} are
followed in the rest of our work.
\section{The three operations' applications to derivatives}
Fix, once and for all, a 4-dimensional manifold $M$, with a spinor
structure $\{\epsilon_{AB},\bar{\epsilon}_{A'B'}\}$\footnote{Here,
we shall take the soldering form $\sigma^a{}_{AA'}$ as the
fundamental variable, with the metric given by
$g^{ab}=\sigma^a{}_{AA'}\sigma^b{}_{BB'}\epsilon^{AB}\bar{\epsilon}^{A'B'}$\cite{Ashtekar}.}.
Following \cite{Sara}, we would like to introduce three convenient
operations: acute-operation, grave-operation, and hat-operation,
each of which means a mapping, from mixed tensor-spinor fields of
rank $(i,j|k,l;m,n)$ to fields of rank
$(i+1,j+1|k+1,l+1;m+1,n+1)$. For example, given a tensor-spinor
field of rank $(1,1|1,1;1,1)$ $\psi^a{}_b{}^{EF'}{}_{GH'}$, then
the three operations are defined as
\begin{equation}
\acute{\psi}^{ac}{}_{bd}{}^{EF'II'}{}_{GH'JJ'}=\frac{1}{4}(\psi^c{}_b{}^{EF'}{}_{GH'}\delta^a{}_d\epsilon^I{}_J\bar{\epsilon}^{I'}{}_{J'}-\psi^a{}_d{}^{EF'}{}_{GH'}\delta^c{}_b\epsilon^I{}_J\bar{\epsilon}^{I'}{}_{J'}),
\end{equation}
\begin{eqnarray}
\grave{\psi}^{ac}{}_{bd}{}^{EF'II'}{}_{GH'JJ'}=&\frac{1}{8}(\psi^a{}_b{}^{IF'}{}_{GH'}\delta^c{}_d\epsilon^E{}_J\bar{\epsilon}^{I'}{}_{J'}+\psi^a{}_b{}^{EI'}{}_{GH'}\delta^c{}_d\epsilon^I{}_J\bar{\epsilon}^{F'}{}_{J'}\nonumber\\
&-\psi^a{}_b{}^{EF'}{}_{JH'}\delta^c{}_d\epsilon^I{}_G\bar{\epsilon}^{I'}{}_{J'}-\psi^a{}_b{}^{EF'}{}_{GJ'}\delta^c{}_d\epsilon^I{}_J\bar{\epsilon}^{I'}{}_{H'}),
\end{eqnarray}
and
\begin{equation}
\hat{\psi}^{ac}{}_{bd}{}^{EF'II'}{}_{GH'JJ'}=\acute{\psi}^{ac}{}_{bd}{}^{EF'II'}{}_{GH'JJ'}+\grave{\psi}^{ac}{}_{bd}{}^{EF'II'}{}_{GH'JJ'}.
\end{equation}
It is easy to generalize these operations to fields of other
ranks; and know that acute-operation on pure spinor fields
vanishes, since there is no tensor index to be replaced.
Similarly, grave-operation on pure tensor fields vanishes;
especially, the three operations on a scalar field all vanish. One
checks these operations: i)are additive, ii)satisfy the Leibnitz
rule under out products, iii)commute with tensor and spinor
contraction, and iv)commute with the complex-conjugation.

When there is no danger of confusion, we shall suppress the
unnecessary indices. For instance, we shall denote
$\psi^a{}_b{}^{EF'}{}_{GH'}$ by $\psi$, and
$\hat{\psi}^{ac}{}_{bd}{}^{EF'II'}{}_{GH'JJ'}$ by
$\hat{\psi}^c{}_d{}^{II'}{}_{JJ'}$.

To show the convenience of introducing these operations, we would
like to present some examples, which are also useful to the rest
of our work. Firstly, given a soldering form $\sigma^b{}_{BB'}$,
let $\nabla_a$ be the unique covariant derivative operator
associated with $\sigma^b{}_{BB'}$, then we have\cite{Wald}
\begin{eqnarray}
(\nabla_a\nabla_b-\nabla_b\nabla_a)\psi&=&-(R_{abc}{}^d\epsilon_I{}^J\bar{\epsilon}_{I'}{}^{J'}\acute{\psi}^c{}_d{}^{II'}{}_{JJ'}+\delta_c{}^dR_{abII'}{}^{JJ'}\grave{\psi}^c{}_d{}^{II'}{}_{JJ'})\nonumber\\
&=&-(R_{abc}{}^d\epsilon_I{}^J\bar{\epsilon}_{I'}{}^{J'}+\delta_c{}^dR_{abII'}{}^{JJ'})\hat{\psi}^c{}_d{}^{II'}{}_{JJ'}.
\end{eqnarray}
Furthermore, let $\xi^a$ be a Killing vector field with respect to
$\sigma^b{}_{BB'}$, then we have\cite{Geroch}
\begin{eqnarray}
\mathcal{D}(\xi)\psi=(\pounds_\xi-\xi^a\nabla_a)\psi&=&-(\nabla_c\xi^d\epsilon_I{}^J\bar{\epsilon}_{I'}{}^{J'}\acute{\psi}^c{}_d{}^{II'}{}_{JJ'}+\delta_c{}^d\nabla_{II'}\xi^{JJ'}\grave{\psi}^c{}_d{}^{II'}{}_{JJ'})\nonumber\\
&=&-(\nabla_c\xi^d\epsilon_C{}^D\bar{\epsilon}_{C'}{}^{D'}+\delta_c{}^d\nabla_{II'}\xi^{JJ'})\hat{\psi}^c{}_d{}^{II'}{}_{JJ'}.\label{Lie}
\end{eqnarray}
Especially, $D(\xi)\psi$ vanishes for all translation Killing
vector fields in a flat spacetime. By
$\nabla_a\nabla_b\xi^c=\xi^dR_{dab}{}^c$\cite{Wald}, we have
\begin{eqnarray}
(\pounds_\xi\nabla_a-\nabla_a\pounds_\xi)\psi&=&\xi^b(\nabla_b\nabla_a-\nabla_a\nabla_b)\psi\nonumber\\
&&+\nabla_a(\nabla_c\xi^d\epsilon_I{}^J\bar{\epsilon}_{I'}{}^{J'}+\delta_c{}^d\nabla_{II'}\xi^{JJ'})]\hat{\psi}^c{}_d{}^{II'}{}_{JJ'}\nonumber\\
&=&-\xi^b(R_{bac}{}^d\epsilon_I{}^J\bar{\epsilon}_{I'}{}^{J'}+\delta_c{}^dR_{baII'}{}^{JJ'})\hat{\psi}^c{}_d{}^{II'}{}_{JJ'}\nonumber\\
&&+\xi^b(R_{bac}{}^d\epsilon_I{}^J\bar{\epsilon}_{I'}{}^{J'}+\delta_c{}^dR_{baII'}{}^{JJ'})\hat{\psi}^c{}_d{}^{II'}{}_{JJ'}\nonumber\\
&=&0,
\end{eqnarray}
which means the Lie-derivative operator via a Killing vector field
commutes with the covariant derivative operator.

Next, given another soldering form $\tilde{\sigma}^b{}_{BB'}$ and
the associated covariant derivative operator $\tilde{\nabla}_a$,
we have\cite{HHZ}
\begin{equation}
(\tilde{\nabla}_a-\nabla_a)\psi=C^d{}_{ac}\epsilon_I{}^J\bar{\epsilon}_{I'}{}^{J'}\acute{\psi}^c{}_d{}^{II'}{}_{JJ'}+\delta_c{}^d\Gamma_{aII'}{}^{JJ'}\grave{\psi}^c{}_d{}^{II'}{}_{JJ'},\label{Variation1}
\end{equation}
where
\begin{equation}
C^d{}_{ac}=\frac{1}{2}\tilde{g}^{db}(\nabla_a\tilde{g}_{cb}+\nabla_c\tilde{g}_{ab}-\nabla_b\tilde{g}_{ac}),\label{Variation2}
\end{equation}
and
\begin{equation}
\Gamma_{aII'}{}^{JJ'}=\frac{1}{2}[(\nabla_a\tilde{\sigma}^b{}_{IA'}+C^b{}_{ac}\tilde{\sigma}^c{}_{IA'})\tilde{\sigma}_b{}^{JA'}\bar{\epsilon}_{I'}{}^{J'}+(\nabla_a\tilde{\sigma}^b{}_{AI'}+C^b{}_{ac}\tilde{\sigma}^c{}_{AI'})\tilde{\sigma}_b{}^{AJ'}\epsilon_I{}^J].\label{Variation3}
\end{equation}
\section{The equivalence of the two energy-momentum tensors}
Start with the lagrangian $\mathcal{L}$ for any mixed
tensor-spinor field $\psi$, which is a local function of the
spinor structure $\{\epsilon_{AB},\bar{\epsilon}_{A'B'}\}$, the
soldering form $\sigma^a{}_{AA'}$, the tensor-spinor field $\psi$
and its first derivative, i.e.,
\begin{equation}
\mathcal{L}_p=\mathcal{L}[\epsilon_{AB}(p),\bar{\epsilon}_{A'B'}(p),\sigma^a{}_{AA'}(p),\psi(p),\nabla_a\psi(p)].
\end{equation}

For a flat spacetime $(M,\eta_{ab})$, let $\xi^a$ be a Killing
vector field, then we have
\begin{equation}
\nabla_a(\xi^a\mathcal{L})=\xi^a\nabla_a\mathcal{L}=\pounds_\xi\mathcal{L}=\frac{\partial
\mathcal{L}}{\partial \psi}\pounds_\xi\psi+\frac{\partial
\mathcal{L}}{\partial \nabla_a\psi}\pounds_\xi\nabla_a\psi,
\end{equation}
where we have used\cite{Geroch}
\begin{equation}
\pounds_\xi\sigma^a{}_{AA'}=0,
\end{equation}
and
\begin{equation}
\pounds_\xi\epsilon_{AB}=0.
\end{equation}
Employing $\pounds_\xi\nabla_a\psi=\nabla_a\pounds_\xi\psi$, and
the equation of motion for $\psi$, i.e.,
\begin{equation}
\frac{\partial \mathcal{L}}{\partial \psi}=\nabla_a\frac{\partial
\mathcal{L}}{\partial \nabla_a\psi},
\end{equation}
we obtain
\begin{equation}
\nabla_a(\frac{\partial \mathcal{L}}{\partial
\nabla_a\psi}\pounds_\xi\psi-\xi^a\mathcal{L})=0.\label{Noether}
\end{equation}
Noting that $D(\xi)\psi$ vanishes for all translation Killing
vector fields, Eq.(\ref{Noether}) implies the canonical
energy-momentum tensor
\begin{equation}
T_\mathcal{C}^{ab}=\frac{\partial \mathcal{L}}{\partial
\nabla_a\psi}\nabla^b\psi-\eta^{ab}\mathcal{L}\label{Can}
\end{equation}
is conserved, i.e., $\nabla_aT_\mathcal{C}^{ab}=0$. Next, let
$\{x^\mu\}$ be the Lorentzian coordinate system, and
$\xi^{\mu\nu}$ be rotation or boost Killing vector fields, i.e.,
\begin{equation}
\xi^{\mu\nu}_b=x^\mu (dx^\nu)_b-x^\nu (dx^\mu)_b,\label{Belin1}
\end{equation}
then Eq.(\ref{Noether}) can be written as
\begin{equation}
\nabla_a[\frac{\partial \mathcal{L}}{\partial
\nabla_a\psi}\mathcal{D}(\xi^{\mu\nu})\psi+T_\mathcal{C}^{ab}\xi^{\mu\nu}_b]=0.\label{Belin2}
\end{equation}
Using $\nabla_aT_\mathcal{C}^{ab}=0$, Eq.(\ref{Belin1}) and
Eq.(\ref{Belin2}) imply
\begin{equation}
T_\mathcal{C}^{[\mu\nu]}=-\frac{1}{2}\nabla_a[\frac{\partial
\mathcal{L}}{\partial
\nabla_a\psi}\mathcal{D}(\xi^{\mu\nu})\psi].\label{Symmetry1}
\end{equation}
Define the tensors
\begin{equation}
N^{abc}=\frac{\partial\mathcal{L}}{\partial\nabla_a\psi}{\cal
D}(\xi^{\mu\nu})\psi(\frac{\partial}{\partial
x^{\mu}})^b(\frac{\partial}{\partial x^{\nu}})^c,\label{Symmetry3}
\end{equation}
which is antisymmetric for the last two indices; and
\begin{equation}
F^{abc}=\frac{1}{2}(N^{abc}-N^{cab}+N^{bca}),\label{Symmetry2}
\end{equation}
which satisfies $F^{cab}=F^{[ca]b}$ and $F^{cab}=F^{c[ab]}$. Then
the conserved and symmetric Belinfante's energy-momentum tensor
can be constructed as
\begin{equation}
T_\mathcal{B}^{ab}=T_\mathcal{C}^{ab}+\nabla_cF^{cab}=T_\mathcal{C}^{(ab)}+\nabla_cN^{(ab)c}.\label{Bel}
\end{equation}
For convenience, we shall write $N^{abc}$ explicitly as
\begin{equation}
N^{abc}=-2\frac{\partial {\cal
 L}}{\partial\nabla_a\psi}(\delta_e{}^{[b}\eta^{c]d}\epsilon_I{}^J\bar{\epsilon}_{I'}{}^{J'}+\delta_e{}^d\sigma^{[b}{}_{II'}\sigma^{c]JJ'})\hat{\psi}^e{}_d{}^{II'}{}_{JJ'},\label{N}
\end{equation}
where Eq.(\ref{Lie}), Eq.(\ref{Belin1}), and Eq.(\ref{Symmetry3})
have been employed.

It is worth noting that although we resort to Killing vector
fields to construct Belinfante's energy momentum tensor; according
to Eq.(\ref{Can}), Eq.(\ref{Bel}) and Eq.(\ref{N}),
$T_\mathcal{B}^{ab}$ itself does not depend on these Killing
vector fields. $T_\mathcal{B}^{ab}$ depends on the spinor
structure, the soldering form, the tensor-spinor field and its
first derivative instead, thus $T_\mathcal{B}^{ab}$ constructed in
the flat spacetime can naturally extend to general curved
spacetime: only replace everywhere the soldering form and the
covariant derivative operator associated with $\eta_{ab}$ by those
associated with $g_{ab}$.

On the other hand, performing the variation of the action
\begin{equation}
S=\int\mathcal{L}\sqrt{-g}\mathbf{d^4x},
\end{equation}
we have
\begin{eqnarray}
\delta S=&&\int[\frac{\partial
\mathcal{L}}{\partial\sigma^a{}_{AA'}}\delta\sigma^a{}_{AA'}+\frac{\partial
\mathcal{L}}{\partial\nabla_a\psi}\delta\nabla_a)\psi-\frac{1}{2}\mathcal{L}g_{ab}\delta
g^{ab}]\sqrt{-g}\mathbf{d^4x}\nonumber\\
=&&\int\frac{1}{2}[(\frac{\partial
\mathcal{L}}{\partial\sigma^a{}_{AA'}}\sigma_{bAA'}-\mathcal{L}g_{ab})\delta
g^{ab}+2\frac{\partial
\mathcal{L}}{\partial\nabla_a\psi}\delta\nabla_a)\psi]\sqrt{-g}\mathbf{d^4x}.
\end{eqnarray}
Here, without loss of generalization, with respect to the
variation of  the metric, the variation of the soldering form has
been gauge-fixed as
\begin{equation}
\delta\sigma^a{}_{AA'}=\frac{1}{2}\sigma_{bAA'}\delta
g^{ab}.\label{Gauge}
\end{equation}
Noting that
\begin{equation}
\acute{\mathcal{L}}^c{}_d{}^{II'}{}_{JJ'}=0=\frac{1}{4}\frac{\partial
\mathcal{L}}{\partial\sigma^a{}_{AA'}}\delta^a{}_d\sigma^c{}_{AA'}\epsilon^I{}_J\bar{\epsilon}^{I'}{}_{J'}+\frac{\partial
\mathcal{L}}{\partial
\psi}\acute{\psi}^c{}_d{}^{II'}{}_{JJ'}+\frac{\partial
\mathcal{L}}{\partial
\nabla_a\psi}[\nabla_a\acute{\psi}^c{}_d{}^{II'}{}_{JJ'}-\frac{1}{4}\delta^c{}_a\nabla_d\psi)\epsilon^I{}_J\bar{\epsilon}^{I'}{}_{J'}],
\end{equation}
we have
\begin{equation}
\frac{\partial
\mathcal{L}}{\partial\sigma^a{}_{AA'}}\sigma_{bAA'}=\frac{\partial\mathcal{L}}{\partial
\nabla^b\psi}\nabla_a\psi-\nabla_c(\frac{\partial\mathcal{L}}{\partial
\nabla_c\psi}\acute{\psi}_{ba}{}^{II'}{}_{JJ'})\epsilon_I{}^J\bar{\epsilon}_{I'}{}^{J'}.
\end{equation}
Therefore, the variation of the action can be written as
\begin{equation}
\delta S=\int\frac{1}{2}[T_{\mathcal{C}(ab)}\delta
g^{ab}-\nabla_c(\frac{\partial\mathcal{L}}{\partial
\nabla_c\psi}\acute{\psi}_{ba}{}^{II'}{}_{JJ'})\epsilon_I{}^J\bar{\epsilon}_{I'}{}^{J'}\delta
g^{ab}+2\frac{\partial
\mathcal{L}}{\partial\nabla_a\psi}\delta\nabla_a)\psi]\sqrt{-g}\mathbf{d^4x}
\end{equation}
According to Eq.(\ref{Variation1}), Eq.(\ref{Variation2}),
Eq.(\ref{Variation3}), and Eq.(\ref{Gauge}), we have
\begin{eqnarray}
\delta\nabla_a)\psi&=&\delta
C^d{}_{ac})\epsilon_I{}^J\bar{\epsilon}_{I'}{}^{J'}\acute{\psi}^c{}_d{}^{II'}{}_{JJ'}+\delta_c{}^d\delta\Gamma_{aII'}{}^{JJ'}\grave{\psi}^c{}_d{}^{II'}{}_{JJ'}\nonumber\\
&=&\frac{1}{2}g_{ec}(g_{af}\nabla^d\delta
g^{ef}-\nabla_a\delta g^{de}-g_{af}\nabla^e\delta g^{df})\epsilon_I{}^J\bar{\epsilon}_{I'}{}^{J'}\acute{\psi}^c{}_d{}^{II'}{}_{JJ'}\nonumber\\
&&+\delta_c{}^dg_{ab}\nabla^e\delta
g^{bf})\sigma_{[fII'}\sigma_{e]}{}^{JJ'}\grave{\psi}^c{}_d{}^{II'}{}_{JJ'}.
\end{eqnarray}
Thus, making partial integrations and employing the Stokes
theorem, we have
\begin{eqnarray}
 \delta S=&&\int\frac{1}{2}[T_{\mathcal{C}(ab)}\delta
g^{ab}-\nabla_c(\frac{\partial\mathcal{L}}{\partial
\nabla_c\psi}\acute{\psi}_{ba}{}^{II'}{}_{JJ'})\epsilon_I{}^J\bar{\epsilon}_{I'}{}^{J'}\delta
g^{ab}+\nabla_c(\frac{\partial\mathcal{L}}{\partial
\nabla_c\psi}\acute{\psi}_{ba}{}^{II'}{}_{JJ'})\epsilon_I{}^J\bar{\epsilon}_{I'}{}^{J'}\delta
g^{ab}\nonumber\\
&&-2\nabla^c(\frac{\partial\mathcal{L}}{\partial
\nabla^b\psi}\acute{\psi}_{[ac]}{}^{II'}{}_{JJ'})\epsilon_I{}^J\bar{\epsilon}_{I'}{}^{J'}\delta
g^{ab}-2\nabla^c(\frac{\partial\mathcal{L}}{\partial
\nabla^a\psi}\sigma_{[bII'}\sigma_{c]}{}^{JJ'}\grave{\psi}^c{}_d{}^{II'}{}_{JJ'})\delta_c{}^d\delta g^{ab}]\sqrt{-g}\mathbf{d^4x}\nonumber\\
=&&\int\frac{1}{2}(T_{\mathcal{C}(ab)}+\nabla^cN_{(ab)c})\delta
g^{ab}\sqrt{-g}\mathbf{d^4x}.
\end{eqnarray}
Obviously, if we define the metric energy-momentum tensor
$T_{\mathcal{M}ab}$ as $\frac{2}{\sqrt{-g}}\frac{\delta S}{\delta
g^{ab}}$, we have
\begin{equation}
T_{\mathcal{B}ab}=T_{\mathcal{M}ab},
\end{equation}
which completes our proof.
\section*{Acknowledgement}
It is my pleasure to acknowledge Prof. R. Geroch for instructive
suggestions and helpful discussions throughout the whole work. I
would also like to give my thanks to Dr. B. Zhou for his
encouragement and discussions. In addition, this work was supported
in part by NSFC(grant 10205002) and NSFC(grant 10373003).

\end{document}